# Charge Distribution Dependency on Gap Thickness of CMS Endcap RPC

**Sung K. Park[1] , Min H. Kang, K. S. Lee**
**On behalf of the CMS Collaboration**

*Korea Detector Laboratory, Dept. of Physics, Korea University, Anam-dong 5-ga, Sungbuk-gu, Seoul, Republic of Korea*
*E-mail:* **sungpark@korea.ac.kr**

ABSTRACT: We report a systematic study of charge distribution dependency of CMS Resistive Plate Chamber (RPC) on gap thickness. Prototypes of double-gap RPCs with six different gap thickness ranging from from 1.0 to 2.0 mm in 0.2-mm steps have been built with 2-mm-thick phenolic high-pressure-laminated plates. The efficiencies of the six gaps are measured as a function of the effective high voltages. We report that the strength of the electric fields of the gap is decreased as the gap thickness is increased. The distributions of charges in six gaps are measured. The space charge effect is seen in the charge distribution at the higher voltages. The logistic function is used to fit the charge distribution data. Smaller charges can be produced within smaller gas gap. But the digitization threshold should be also lowered to utilize these smaller charges.

---

[1] Corresponding author.

# Contents

## 1. Introduction

As muon trigger detector in CMS experiment, the resistive plate chambers (RPCs) have been demonstrating its capability with good trigger performance in LHC environments. The CMS RPC trigger system can be divided into two parts according to the RPC locations: the central region with six trigger stations and forward Endcap region (RE) with four stations of RE1-RE4 [1].

Within each RE station, there are three concentric sections which divide the RE in terms of pseudorapidities. Until now RE1/2, RE1/3, RE2/2, RE2/3, RE3/2, RE3/3, RE4/2 and RE4/3 are installed and become RPC trigger system which makes pseudorapidity coverage of the RPC trigger up to $|\eta| < 1.6$ [2-3].

The plan has been to install all the remaining RE1/1, RE2/1, RE3/1 and RE4/1 for the full trigger coverage. The current CMS plan is to use GEM technology for RE1/1 but the conventional RPC technology seems applicable to the remaining sections of CMS RPCs. The current R&D for the CMS RPCs is to find a detector solution for the future CMS RPCs in high backgrounds. For the PHASE II upgrade, we need more RPCs in RE, especially RE3/1 and RE4/1, covering in $1.6<|\eta|<2.1$ by 2023.

The CMS RPCs in high backgrounds should have a higher rate capability. The lower resistivity of the electrode contributes to the higher rate capability since the rate capability is inversely proportional to the resistivity of the electrodes. Reduction of the avalanche charges is also a key to enhance the higher rate capability. One way to obtain smaller avalanche charges is to reduce the gas volume by reducing gas gap thickness. The smaller charges certainly reduce the probability of aging due to the high-rate background guaranteeing the longevity of the RPCs. But many advantages of the smaller charges can be only achieved by a lower digitization threshold.

The main theme of this paper is to show that both smaller thickness of RPC gaps and low digitization threshold are required to utilize the advantages of the smaller avalanche charges. But lowering the digitization threshold requires a new development of more sensitive front-end electronics compared to the current ones that have been used for the current CMS-RPC operation. Therefore, we concentrate rather on the study of the charge distribution dependency on gap thickness.

## 2. Efficiency in variation of gap thickness

Six RPCs with different gap thickness varying in 0.2-mm steps from 1.0 to 2.0 mm are built to study charge dependency on the gap thickness as shown in Figure 1. We measured the efficiencies and cluster sizes of cosmic muons for the six RPCs with the six different gap thicknesses as a function of high voltages.

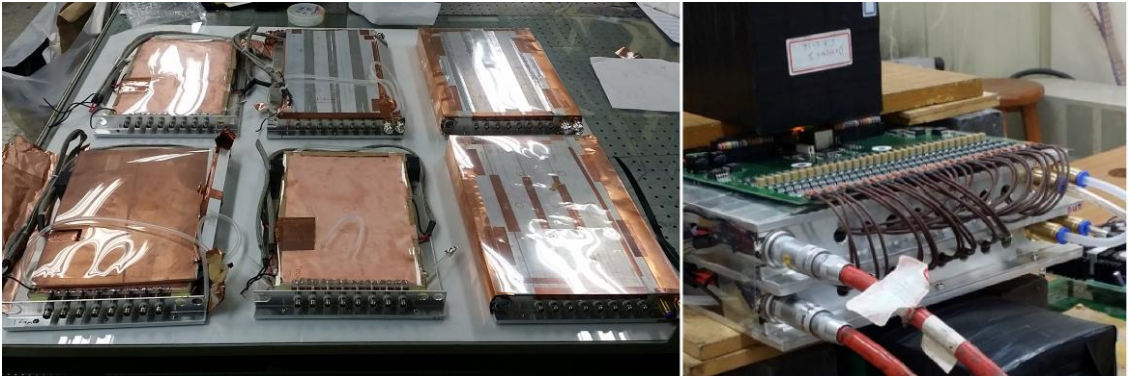

Figure 1. Six RPCs with different gap thickness from 1.0 to 2.0 mm in a step of 0.2 mm.

The applied high voltages are converted to the effective high voltages (H.V) under the standard conditions of $P = 1013$ hPa and $T = 293$K. Figure 2 shows the efficiencies of the cosmic muons measured for 1.0-mm-thick (left) and 2.0-mm-thick (right) double-gap RPCs as a function of H.V. The values of H.V yielding efficiencies of 95% and 50% measured for the six different RPCs are summarized in Table 1.

| Gap thickness (mm) | 1.0 | 1.2 | 1.4 | 1.6 | 1.8 | 2.0 |
|---|---|---|---|---|---|---|
| $H.V_{\varepsilon=0.50}$ (kV) | 5.38 | 6.16 | 6.94 | 7.71 | 8.47 | 9.17 |
| $H.V_{\varepsilon=0.95}$ (kV) | 5.66 | 6.50 | 7.21 | 7.99 | 8.78 | 9.39 |

Table 1. H.Vs measured at the efficiencies of 95% and 50% for the six different RPCs.

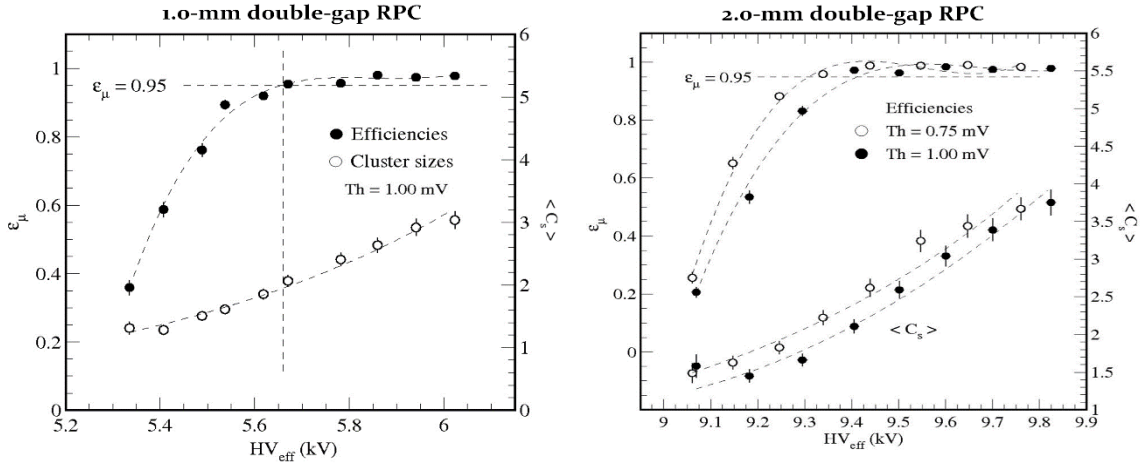

Figure 2. Efficiencies measured for 1.0-mm-thick (left) and 2.0-mm-thick (right) double-gap RPCs as a function of H.V.

As we reduce the gap thickness from 2.0 mm that has been the CMS standard gap thickness to 1.0 mm in a step of 0.2 mm, the values of H.V at 95% and 50% of efficiencies are decreased at lower than a linear rate as shown in Figure 3. But the relationship between the gap thickness and effective high voltages is expected to be a linear relationship.

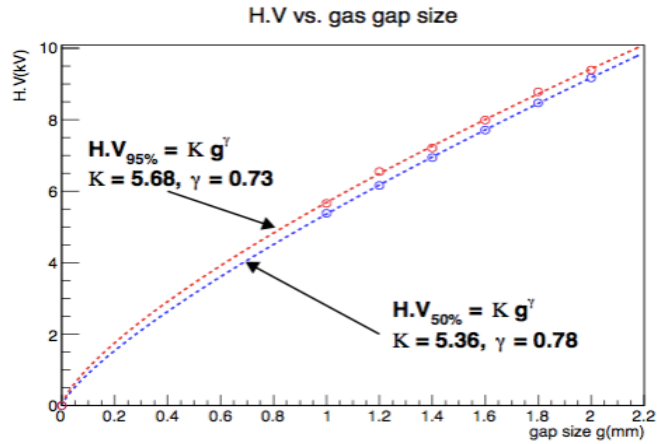

Figure 3. Values of HV$_{\varepsilon=0.50}$ (blue) and HV$_{\varepsilon=0.95}$ (red) as a function of the gap thickness.

We assumed gap-thickness-dependent high voltage, H.V = $K g^\gamma$ for the effect of gap thickness on H.V. Here, $g$ is the gap thickness while $K$ and $\gamma$ are the fitting parameters. For the values of two $K$ and $\gamma$ we fit the data shown in Table 1 with the function as shown in Figure 3 [4].

For the linear relationship between the gap thickness and the H.V, the value of $\gamma$ is 1. But we obtained $\gamma$ to be 0.73 and 0.78 for 95% and 50% of efficiency respectively. When we plot the electric field of the gap in terms of the gap thickness it becomes more obvious that the strength of the electric field is decreasing as the gap thickness is increased. Summary of six gap thickness and their electric fields is listed in Figure 4.

We understand it in terms of the first Townsend coefficient where the avalanche amplification grows by the factor $e^{\alpha g}$, where α is the first Townsend coefficient and g is the gap thickness. Our operation mode is the avalanche mode in which our signal size is above the threshold and below the streamer signal. It means that our gas amplification is fixed within these ranges. As g is increased, the α should be decreased and the strength of the electric field also should be decreased as shown in Figure 4.

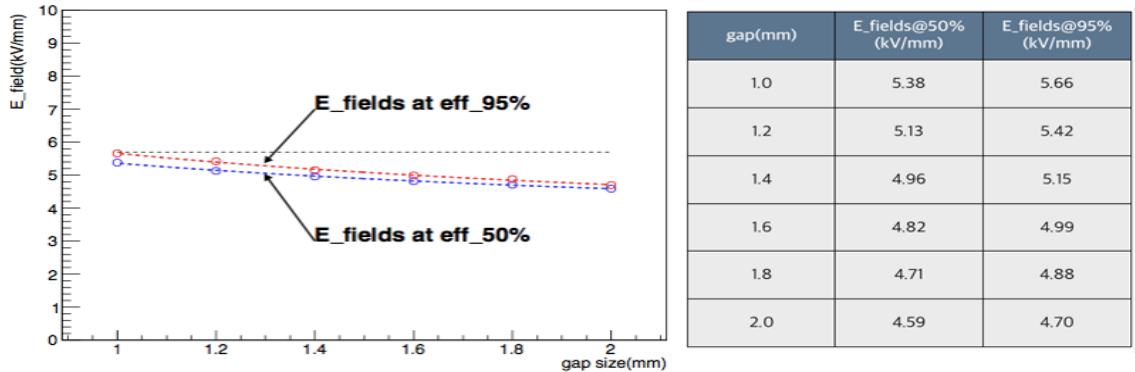

Figure 4. Gap thickness vs. electric fields at 50% and 95% efficiency in avalanche mode. Summary of electric fields of six different gap thickness (right).

## 3. Charge distribution in variation of gap thickness

We also measured the charges in six different gap thickness. Charge measurements of 2.0 mm and 1.0 mm gap thickness are plotted at various H.V in Figure 5. For 2.0 mm gap, at the lower

H.V of 8.96 kV in Figure 5, the charge distribution is mostly populated near zero. As H.V increases the small charges distribution which was mostly populated around zero moves horizontally with its distribution shape remaining intact as shown for 10.02 kV in Figure 5. The small charges are difficult to be found at the higher voltages. Since small charges are produced near the anode, it means that at the higher voltages, the space charges affect near the anode preventing smaller charges from being produced.

This trends can also be found from the charge distribution of 1.0 mm gap in Figure 5. For 1.0 mm gap, the charge distribution is found in Figure 6. As in the case of 2.0 mm, the charge distribution is most popular near the zero at the H.V of 5.27 kV. But at 6.14 kV the smaller charges near the zero diminish significantly. Four other remaining gaps also show the same trend of charge distribution as seen from 2.0 mm and 1.0 mm. From these observation, we can conclude that the smaller charges are not produced due to the space charge effect near the anode at the higher voltages.

We also plot the mean charge together with efficiency curve in terms of the H.V in Figure 6. The charges plotted in logarithm shows that amount of charge collected by pickup strip is proportional to the H.V. But the proportionality slop changes from faster rising to slower rising at around 9.4 (5.4) kV for 2.0 (1.0) mm gap thickness as shown in Figure 5.

The charge distributions from all six gaps are similar in this pattern to each other. But one noticeable pattern is that the transition from faster rising slop to slower rising slop occurs at around the H.V of 95% efficiency for 2.0 mm gap but the transition started earlier at about 300 volts below the H.V of 95% efficiency for 1.0 mm gap. The transition from faster rising to slower rising at higher voltage as shown in Figure 6 and the space charge effect near the anode as shown in Figure 5 at the higher voltages lead us to think about the saturation near the anode.

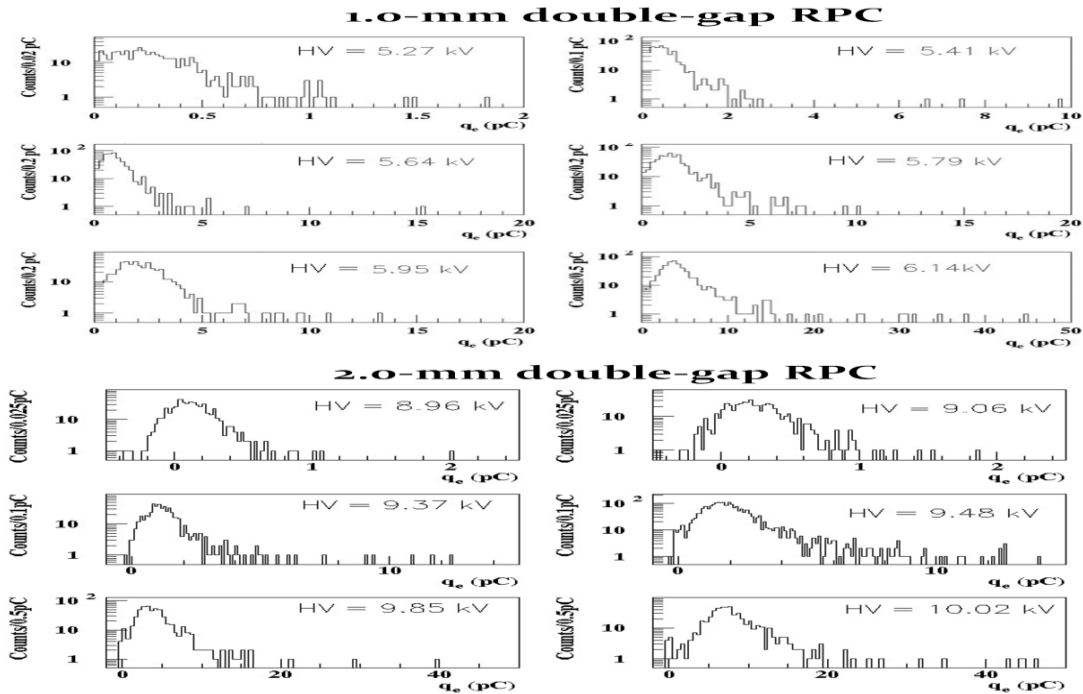

Figure 5. Two columns of the charge distribution at various H.V of 1.0 mm gap (top) and 2.0 mm gap (bottom).

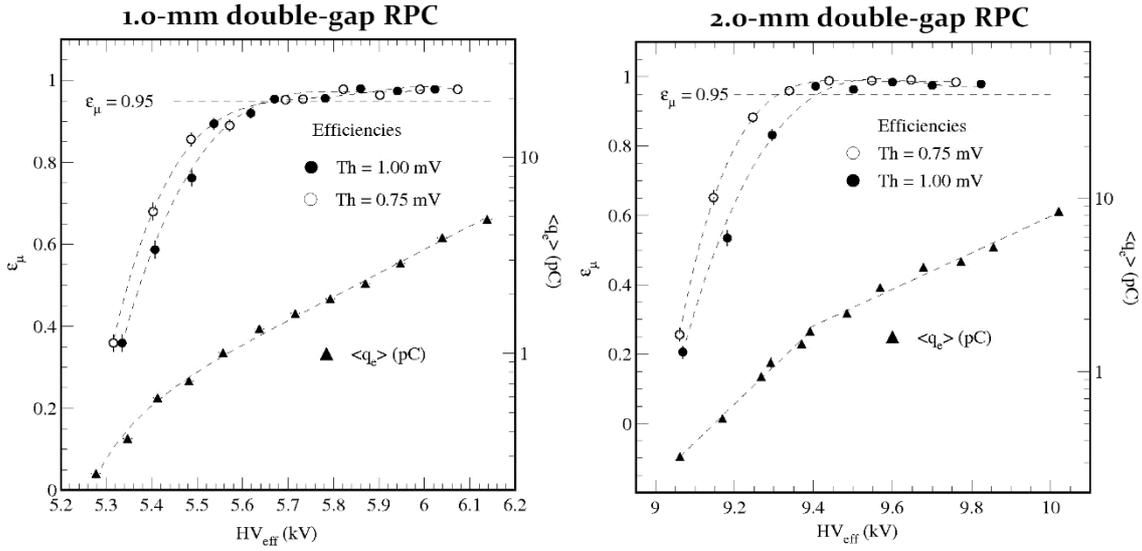

Figure 6. Mean charge and efficiency of 1.0 mm gap (left) and 2.0 mm gap (right).

## 4. Charge distribution models

All of the mean charge distributions of six gaps at the various H.Vs are plotted in Figure 7. The avalanche models of the exponential charge growth doesn't consider the presence of the space charges. This intuitive exponential growth can't explain our charge distribution data. Therefore we need some models to understand these charge distributions of the different gaps. At least, there are two models: one is simulating from the first principles [5] and the other is describing the charge saturation within a logistic model [6]. The logistic function provides simple forms to fit the charge distributions.

Mean charge distributions of 2.0 mm gap and 1.0 mm gap are fitted with the logistic function and the fit results are plotted in Figure 8 where $v_0$ is the H.V at the half of the saturation value, α is a growth rate, K is the saturation value and v is the H.V.

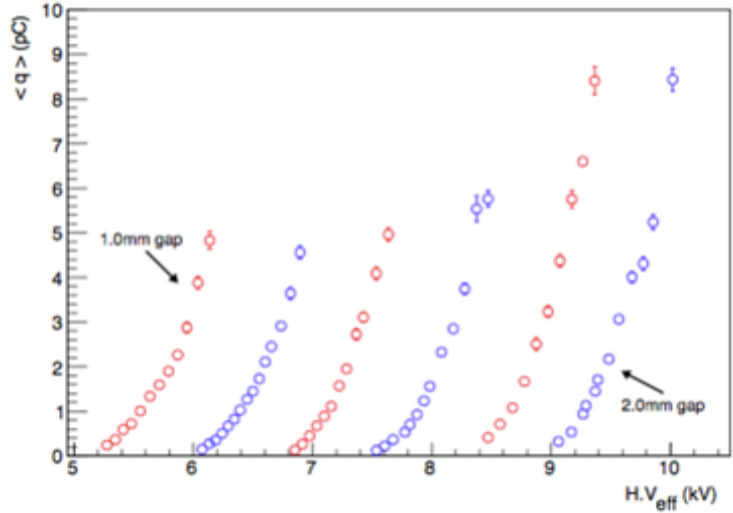

Figure 7. Mean charge distributions of all six gaps at various H.Vs.

The logistic function describes the overall mean charge distribution as a function of H.V reasonably well as shown in Figure 8. From the results of the fit summarized in Table 2, we noticed that $v_0$ in the logistic function is very close to the beginning of 95% of efficiency, at which is already saturated with space charges.

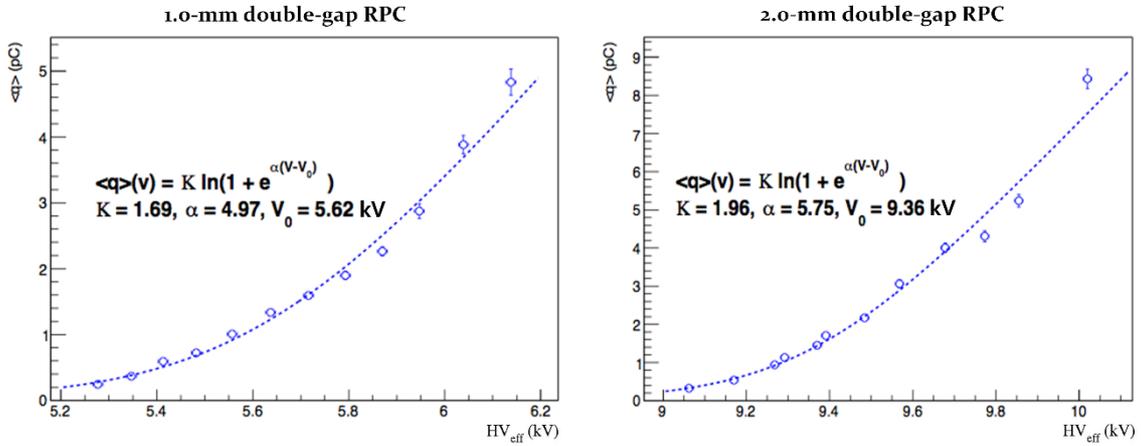

Figure 8. Mean charge distribution of 1.0 mm gap (left) and 2.0 mm gap (right) are fitted with logistic function's cumulative.

The transition points from faster rising to slower rising for all six gaps are related to the saturation due to space charges near anode. But the logistic model doesn't tell us about where the transition point occurs. In future, a more comprehensive model can hopefully describe not only the mean charge distribution but also the location of the saturation due to the space charge.

## 5. Digitization threshold limiting the charge distribution

When the gap thickness is reduced from 2.0 mm to 1.0 mm, the amount of charge reduced is 18% of the 1.66 pC as summarized in Table 2. It is because the digitization threshold value was set too high to utilize the smaller charges produced in the smaller gas gaps. One scenario is that if the digitization threshold values could have been lowered much below the current value, we might have obtained much smaller charge of around 0.5 pC than 1.66 pC which was obtained with the threshold value of 1.0 mV. The charge distributions of 1.0 mm gap in Figure 6 supports this scenario.

## 6. Conclusions

A systematic study of charge distribution dependency on gap thickness has been carried out. The efficiencies of the six gaps are measured as a function of the H.V. The strength of the electric fields of the six gaps is decreased as the gap thickness is increased.

The distributions of the charges in six gaps are measured. The space charge effect is seen in the charge distribution at the higher voltages. The logistic function is used to fit the charge distribution of the six gaps.

As gap thickness is decreased, the amount of the pickup charge is decreased. But the reduction of the charge is not so much as expected. This is due to the higher digitization threshold.

A smaller avalanche charge can be utilized only by the combination of the smaller gap thickness and the lower digitization threshold.

| Gap (mm) | H.V_95%(kV) Th(1.0mV) | <q_e>(pC) | V_0(kV) in logistic fun. |
|---|---|---|---|
| 2.0 | 9.39 | 1.658+/-0.108 | 9.36 |
| 1.8 | 8.77 | 1.621+/-0.072 | 8.80 |
| 1.6 | 7.99 | 1.607+/-0.080 | 7.91 |
| 1.4 | 7.21 | 1.473+/-0.069 | 7.05 |
| 1.2 | 6.50 | 1.448+/-0.049 (94%) | 6.36 |
| 1.0 | 5.66 | 1.423+/-0.079 | 5.62 |

Table 2. Summary of mean charge of six gaps as a function of H.V. Mean charges are measured at 95% efficiency except 1.2 mm gap where 94% is used.

## Acknowledgments

This study was supported by the National Research Foundation of Korea (grant number NRF-2016K1A3A1A25003501). Most of all, we would like to give our special thanks to all team members dedicated the beam test and to the CERN EN and EP departments for the facility infrastructure support.